\documentclass{PoS}
\pdfoutput=1 

\usepackage{graphics}
\usepackage{amsmath} 
\usepackage{amsfonts}
\usepackage{mathtools}
\usepackage{pstricks}
\usepackage[final]{pdfpages} 
\usepackage{ifpdf} 
\RequirePackage{ifpdf} 

\usepackage{color}
\definecolor{urlblue}{rgb}{0.2,0.4,0.7}
\definecolor{citegreen}{rgb}{0,0.6,0.2}
\definecolor{linkred}{rgb}{0.9,0.2,0.1}

\usepackage{caption} 
\usepackage{subcaption} 
\usepackage{autobreak}
\usepackage{marginnote}
\usepackage{multirow}

\usepackage{etoolbox} 
\usepackage{fixmath}
\usepackage{psfrag}
\usepackage{cases}
\usepackage{slashed}

\usepackage{notoccite} 

\newcommand{\als}{\ensuremath{\alpha_s}}

\newcommand{\mur}{\ensuremath{\mu_R}}
\newcommand{\amp}{\ensuremath{\mathcal{M}}}

\def\nn{\nonumber}

\def\be{\begin{equation}}
\def\ee{\end{equation}}
\def\bea{\begin{eqnarray}}
\def\eea{\end{eqnarray}}
\def\nn{\nonumber}

\newcommand{\secdec}{\textsc{SecDec}{}}
\newcommand{\pysecdec}{py\secdec}
\newcommand{\gosam}{\textsc{GoSam}{}}
\newcommand{\powhegbox}{\texttt{POWHEG-BOX-V2}{}}

\newcommand{\mgg}{m_{\gamma\gamma}}
\newcommand{\mt}{m_t}

\title{\vspace*{-5.15em}
\mbox{}\hfill \mbox{\small\sc MPP-2019-254}\\
\vspace*{6em}
Photon pair production in gluon fusion: Top quark effects around threshold}

\ShortTitle{$g\,g \rightarrow \gamma\, \gamma$}

\author{\speaker{Long Chen} \thanks{L. Chen, G.~Heinrich and M.~Kerner thank the organisers of RADCOR 2019.} \\
        Max Planck Institute for Physics, F\"ohringer Ring 6, 80805 M\"unchen, Germany\\
        E-mail: \email{longchen@mpp.mpg.de}}

\author{Gudrun Heinrich\\
        Max Planck Institute for Physics, F\"ohringer Ring 6, 80805 M\"unchen, Germany\\
        E-mail: \email{gudrun@mpp.mpg.de}}

\author{Stephan Jahn\\
        Max Planck Institute for Physics, F\"ohringer Ring 6, 80805 M\"unchen, Germany\\
        E-mail: \email{sjahn@mpp.mpg.de}}

\author{Stephen P.~Jones\\
        Theoretical Physics Department, CERN, Geneva, Switzerland\\
        E-mail: \email{s.jones@cern.ch}}

\author{Matthias Kerner\\
        Physik-Institut, Universit{\"a}t Z{\"u}rich, Winterthurerstrasse 190, 8057 Z{\"u}rich, Switzerland\\
        E-mail: \email{mkerner@physik.uzh.ch}}

\author{Johannes Schlenk\\
        Theory Group LTP, Paul Scherrer Institut, CH-5232 Villigen PSI, Switzerland\\
        E-mail: \email{johannes.schlenk@psi.ch}}

\author{Hiroshi Yokoya\\
        Quantum Universe Center, KIAS, Seoul 02455, Korea\\
        E-mail: \email{hyokoya@kias.re.kr}}

\abstract{
We present a calculation of the NLO QCD corrections to the loop-induced production of a photon pair through gluon fusion, 
including massive top quarks at two loops.
The two-loop virtual amplitudes are computed via projecting onto a set of Lorentz structures related to 
linear polarisation states of external gauge bosons. All two-loop master integrals dependent on the top quark mass 
are evaluated numerically. 
Matching the fixed order NLO result to a calculation with threshold resummation, we obtain an accurate description of 
the diphoton invariant mass spectrum around the threshold of top quark pair production.
We analyse how the distribution of the diphoton invariant mass is affected by the top quark loop-induced corrections, 
especially around the top quark pair production threshold.
}

\FullConference{14th International Symposium on Radiative Corrections (RADCOR2019)\\ 
                9-13 September 2019,~ Palais des Papes, Avignon, France}

\begin{document}
\allowdisplaybreaks[4]

\maketitle

\section{Introduction}

Diphoton production is an important process at hadron colliders for phenomenological studies 
from both the experimental and theoretical side.
Most prominently, the diphoton final state served as one of the key discovery channels 
for the Higgs boson, even though the branching ratio 
of the diphoton decay channel is just about a few per-mille. 
Up to now it is still the channel with the highest mass resolution.
As being experimentally very clean, a precise knowledge about the diphoton spectrum 
at hadron colliders is desirable both from the point of view of examining the dynamics of the Standard Model, 
and for searching possible manifestations of Beyond-Standard-Model physics.

Another interesting aspect of diphoton production is the possibility of measuring the top quark mass 
via the characteristic threshold effects manifest in the diphoton invariant mass spectrum around 
the top quark pair production threshold~\cite{Kawabata:2016aya}.
As discussed in ref.~\cite{Kawabata:2016aya}, while current LHC measurements are not yet able to provide 
the necessary statistics for such an analysis, the feasibility at a future hadron collider 
is worth investigating.
To this end a more precise theoretical computation of the diphoton invariant mass spectrum, 
especially around the top quark pair production threshold, is needed, which is the focus of this work.
~\\

To categorize different contributing channels to the $pp\to \gamma\gamma$ cross section, 
let us briefly recapitulate previous work on this subject. 
Direct diphoton production\footnote{We denote by ``direct photons'' the photons produced directly in the hard scattering process, 
as opposed to photons originating from a hadron fragmentation process.} in hadronic collisions proceeds  
via $q\bar{q}\to \gamma\gamma$ at the leading order (LO), of order $\als^0$ in the QCD coupling. 
The next-to-leading order (NLO) QCD corrections, including fragmentation contributions at NLO, 
were implemented in the public program {\tt Diphox}~\cite{Binoth:1999qq}.
The next-to-next-to-leading order (NNLO) QCD corrections to $pp\to \gamma\gamma$ were first completed in ref.~\cite{Catani:2011qz}, 
including the loop-induced $gg\to\gamma\gamma$ contribution (at order $\als^2$) with massless quark loops.

The loop-induced subprocess $gg\to\gamma\gamma$~\cite{Dicus:1987fk} enters starting from NNLO in QCD (order $\als^2$). Even though the $gg\to\gamma\gamma$ contribution 
to $pp\to \gamma\gamma$ is of higher orders in $\als$, it is comparable in size to that of the LO process 
$q\bar{q}\to \gamma\gamma$ at the LHC, due to the large gluon luminosity.
NLO QCD corrections to the gluon fusion channel with massless quarks, i.e.~order $\als^3$ corrections, 
have been first calculated in refs.~\cite{Bern:2001df,Bern:2002jx} and implemented in  
2$\gamma${\tt MC}~\cite{Bern:2002jx} as well as in MCFM~\cite{Campbell:2016yrh}. 
Diphoton production at NNLO with massless quarks is also available in {\sc Matrix}~\cite{Grazzini:2017mhc}.

Very recently, the NLO QCD corrections to the gluon fusion channel including massive top quark loops have become available~\cite{Maltoni:2018zvp}.
In this work, we first provide an independent calculation of the QCD corrections to the process $gg\to\gamma\gamma$ 
including massive top quark loops, and then we combine our fixed order result with a threshold-resummation improved calculation as advocated in ref.~\cite{Kawabata:2016aya}.

\section{Building blocks of the fixed-order calculation}
\label{sec:calculation}


We consider the following scattering process,
\begin{align}\label{kinematicassignment}
g(p_1, \lambda_1, a_1) + g(p_2, \lambda_2, a_2) \to \gamma(p_3,\lambda_3) + \gamma(p_4,\lambda_4) ,
\end{align}
with on-shell conditions $p_j^2 = 0, j=1,...,4$.  
The helicities $\lambda_i$ of the external particles are defined by taking the momenta of the gluons 
$p_1$ and $p_2$ (with color indices $a_1$ and $a_2$, respectively) 
as incoming and the momenta of the photons $p_3$ and $p_4$ as outgoing.
The Mandelstam invariants associated with eq.~(\ref{kinematicassignment}) are defined as 
$\, s = \left(p_1 + p_2 \right)^2 ,\, t = \left(p_2 - p_3 \right)^2 ,\, u = \left(p_1 - p_3 \right)^2$.

\subsection*{Projection operators}

Pulling out the polarisation vectors $\varepsilon_{\lambda_i}^{\mu_i}$ of external gauge bosons from the amplitude $\amp$ describing the process eq.~(\ref{kinematicassignment}), one defines the tensor amplitude $\amp_{\mu_1\mu_2\mu_3\mu_4}$ by
\begin{equation} \label{eq:ggyyamplitudes}
\amp{} = \varepsilon_{\lambda_1}^{\mu_1}(p_1)\,\varepsilon_{\lambda_2}^{\mu_2}(p_2)\,\varepsilon_{\lambda_3}^{\mu_3,\star}(p_3)\,\varepsilon_{\lambda_4}^{\mu_4,\star}(p_4)\,\amp_{\mu_1\mu_2\mu_3\mu_4}(p_1,p_2,p_3,p_4)\,.
\end{equation}
We compute $\amp_{\mu_1\mu_2\mu_3\mu_4}$ through projection onto a set of Lorentz structures 
related to linear polarisation states of the external gauge bosons, with the corresponding 
D-dimensional projection operators constructed following the prescription proposed 
in ref.~\cite{Chen:2019wyb}.\footnote{This approach has been applied recently in the calculation of ref.~\cite{Ahmed:2019udm}.} These linear polarisation projectors are based on the momentum basis representations of external polarisation vectors (for both bosons and fermions, massless or massive), and all their open Lorentz indices are by definition taken to be D-dimensional to facilitate a uniform projection with just one dimensionality D=$g_{~\mu}^{\mu}$.

For the process in question, we introduce two linear polarisation states $\varepsilon^{\mu}_X$ and $\varepsilon^{\mu}_T$ lying within the scattering plane determined by the three linearly independent external momenta $\{ p_1, p_2, p_3 \}$, and transversal to $p_1$ and $p_3$ respectively. 
In addition, a third linear polarisation state vector $\varepsilon^{\mu}_Y$, orthogonal to $p_1, p_2,$ and $p_3$, 
is constructed with the help of the Levi-Civita symbol. 
To determine the momentum basis representations for 
$\varepsilon^{\mu}_{X/T}$, we first write down a Lorentz covariant ansatz and then solve the orthogonality and normalisation conditions of linear polarisation state vectors for the linear decomposition coefficients. Once we establish a definite Lorentz covariant decomposition form 
in 4 dimensions solely in terms of external momenta and kinematic invariants, 
this form is declared as the definition of the corresponding polarisation state vector in D dimensions.

Applied to the scattering process \eqref{kinematicassignment}, this construction leads to eight projectors  
\begin{equation} \label{eq:LPprojectors}
    \varepsilon_{[X,Y]}^{\mu_1} \varepsilon_{[X,Y]}^{\mu_2} \varepsilon_{[T,Y]}^{\mu_3} \varepsilon_{[T,Y]}^{\mu_4},
\end{equation}
where the square bracket $[\cdot{},\cdot{}]$ in the subscripts means either entry, 
and where only the combinations containing an even number of $\varepsilon_Y$ are considered.
This is simply because \eqref{kinematicassignment} is a P-even 2-to-2 scattering process.
Let us emphasize that, in order to end up with an unambiguous form of projectors to be used in D dimensions, 
all pairs of Levi-Civita tensors should be contracted first (as explained in ref.~\cite{Chen:2019wyb}) 
{\em before} being used for the projection of the amplitude.
In this way the aforementioned projectors are expressed solely in terms of external momenta and metric tensors 
whose open Lorentz indices are all set to be D-dimensional. 
We remark that since all projectors thus constructed obey all defining physical constraints, 
the index contraction between \eqref{eq:LPprojectors} and the tensor amplitude $\amp_{\mu_1\mu_2\mu_3\mu_4}$
is always done with the spacetime metric tensor $g_{\mu\nu}$ (rather than the physical polarisation sum rule). 
The usual helicity amplitudes can be composed using the relations between circular and linear polarisation state, 
e.g.~$\varepsilon_{\pm}^\mu(p_1) = \frac{1}{\sqrt{2}} \left( \varepsilon_X^\mu \pm i \varepsilon_Y^\mu \right)$.

\subsection*{Numerical evaluation of amplitudes}

With the linear polarisation projectors defined in \eqref{eq:LPprojectors}, 
we re-computed the LO amplitudes for the process \eqref{kinematicassignment} analytically, 
with both  massless and massive quark loops.
These expressions were implemented in our computational setup for the NLO QCD corrections, 
which we describe below.

The bare scattering amplitudes of the process \eqref{kinematicassignment} beyond LO contain poles 
in the dimensional regulator $\epsilon \equiv (4-D)/2$ arising from ultraviolet (UV) 
as well as soft and/or collinear (IR) regions of the loop momenta. 
In our computation, we renormalise the UV divergences using the $\overline{\text{MS}}$ scheme, 
except for the top quark contribution which is renormalised on-shell. 
For details on the UV renormalisation, please refer to ref.~\cite{Chen:2019fla}.
To deal with the intermediate IR divergences, we employ the FKS subtraction approach~\cite{Frixione:1995ms},
as implemented in the \powhegbox~framework~\cite{Nason:2004rx,Frixione:2007vw,Alioli:2010xd}.
In practice, we need to supply only the finite part of the born-virtual interference, 
under a specific definition~\cite{Alioli:2010xd} in order to combine it with the FKS-subtracted 
real radiation generated within the \gosam/\powhegbox~framework.

\begin{figure}[htb]
\centering
\begin{subfigure}[b]{0.24\textwidth}
  \centering
  \includegraphics[width=0.51\textwidth]{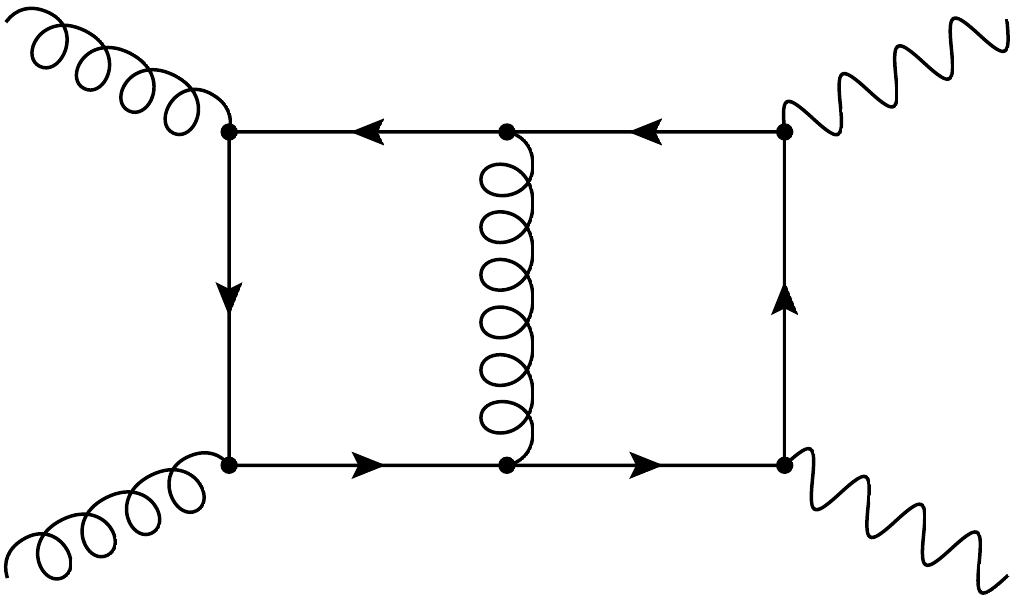} 
\end{subfigure}
\begin{subfigure}[b]{0.24\textwidth}
  \centering
  \includegraphics[width=0.51\textwidth]{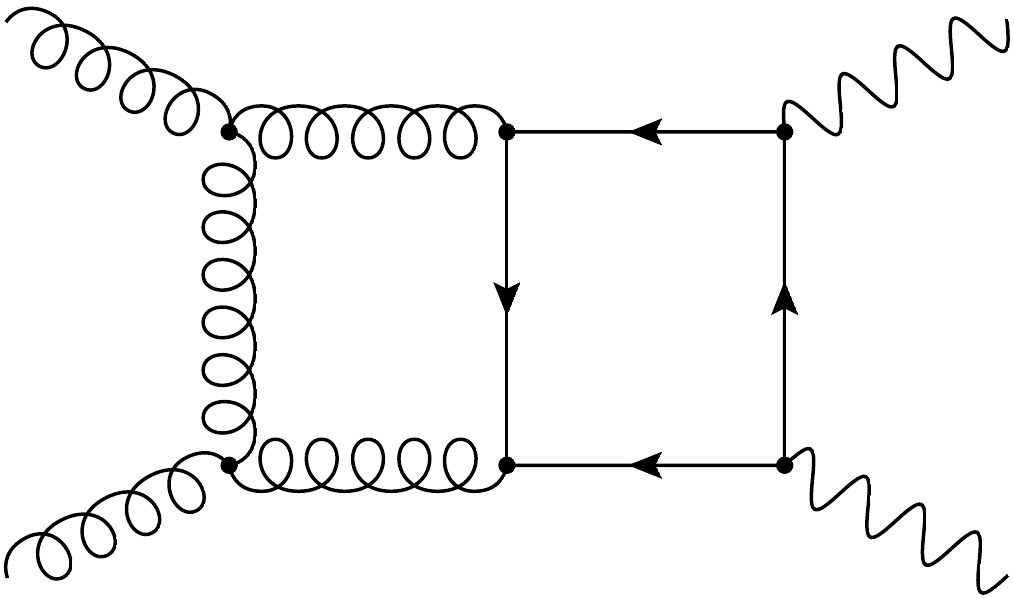} 
\end{subfigure}
\vspace{1em}
\begin{subfigure}[b]{0.24\textwidth}
  \centering
  \includegraphics[width=0.51\textwidth]{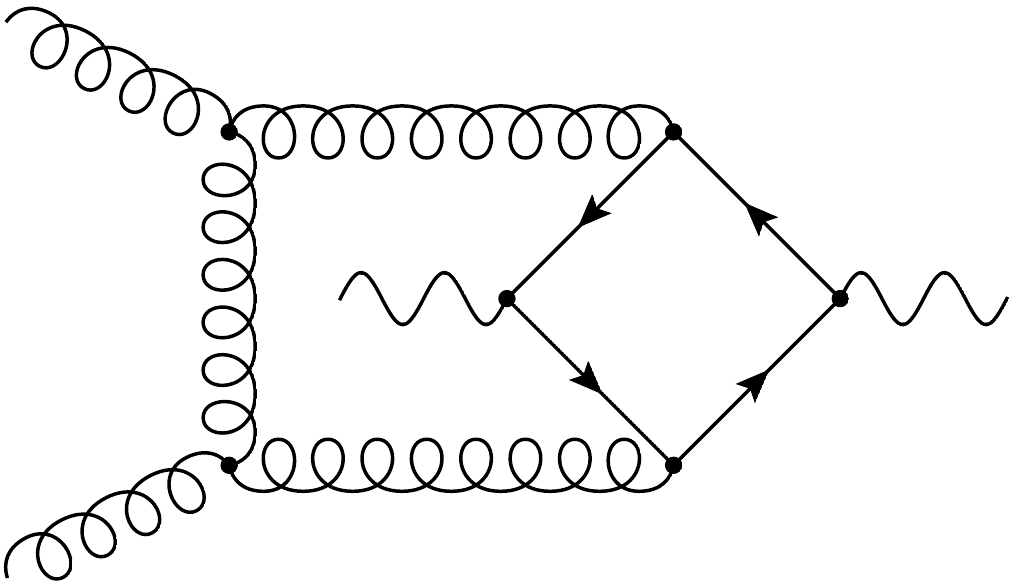} 
\end{subfigure}
\begin{subfigure}[b]{0.24\textwidth}
  \centering
  \includegraphics[width=0.51\textwidth]{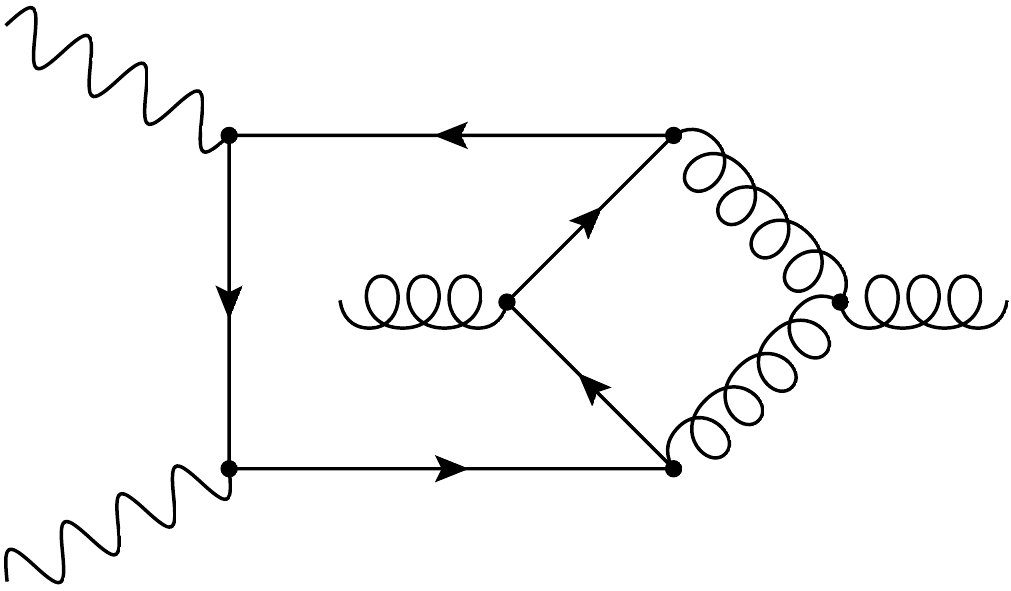} 
\end{subfigure}
\caption{Examples of diagrams contributing to the virtual corrections.}
\label{fig:2l_diagrams}
\end{figure}

For the two-loop QCD diagrams contributing to our scattering process, 
there is a complete separation of quark flavors due to the color algebra and Furry's theorem, 
with samples shown in fig.~\ref{fig:2l_diagrams}.
We obtain the two-loop amplitude with the multi-loop extension of 
the program~\gosam{}~\cite{Jones:2016bci} where {\sc Reduze}\,2~\cite{vonManteuffel:2012np} 
is employed for the reduction to master integrals. 
In particular, each of the linearly polarised amplitudes projected out using \eqref{eq:LPprojectors} 
is eventually expressed as a linear combination of 39 massless 
integrals and 171 integrals that depend on the top quark mass, distributed into three integral families.
All massless two-loop master integrals involved are known analytically~\cite{Bern:2001df,Binoth:2002xg,Argeri:2014qva},
and we have implemented the analytic expressions into our code.   
Regarding the two-loop massive integrals, which are not yet fully known analytically,
we first rotate to an integral basis consisting partly of quasi-finite loop integrals~\cite{vonManteuffel:2014qoa}. 
Our integral basis is chosen such that the second Symanzik polynomial, $\mathcal{F}$, appearing in the Feynman  parametric representation of each of the integrals is raised to a power, $n$, where $|n| \le 1$ in the limit $\epsilon \rightarrow 0$. This choice improves the numerical stability of our calculation near to the $t\bar{t}$ threshold, where the $\mathcal{F}$ polynomial can vanish. All these massive integrals are evaluated numerically using \pysecdec~\cite{Borowka:2017idc,Borowka:2018goh}.

The phase-space integration of the virtual interferences is achieved by reweighting unweighted Born events.
The accuracy goal imposed on the numerical evaluation of the virtual
two-loop amplitudes in the linear polarisation basis in \pysecdec{} is 1 per-mille on 
both the relative and the absolute error.
We have collected 6898 phase space points out of which 862 points fall into the diphoton invariant mass
window $\mgg \in \left[330,\, 360\right]$ GeV. We further have calculated the amplitudes for 2578 more
points restricted to the threshold region.
~\\

The real radiation matrix elements are calculated using the interface~\cite{Luisoni:2013cuh}
between \gosam~\cite{Cullen:2011ac,Cullen:2014yla} and the 
\powhegbox~\cite{Nason:2004rx,Frixione:2007vw,Alioli:2010xd}, modified accordingly to 
compute the real radiation corrections to loop-induced Born amplitudes. 
Only real radiation contributions where both photons couple to a closed quark loop are included.
We also include the $q\bar{q}$ initiated diagrams which contain a closed quark loop, 
even though their contribution is numerically very small.

\section{Treatment of the threshold region}
\label{sec:threshold}

When the $t\bar{t}$ pair inside the loop is produced close to the threshold, the Coulomb interactions between the non-relativistically moving top quarks could lead to divergences in a fixed order calculation.  
To overcome this issue and correctly describe diphoton production around the $t\bar{t}$ threshold,  
we employ the non-relativistic QCD (NRQCD)~\cite{Caswell:1985ui,Bodwin:1994jh,Pineda:1997bj,Beneke:1997zp}, 
which is an effective field theory designed to describe non-relativistic heavy quark-antiquark systems 
in the threshold region.~\\

To the order which we consider here, the amplitude $\amp{}$ can be expressed as a coherent 
sum of light quark loop contributions and the top quark loop contribution,
\begin{align}
\amp{}(p_i,\lambda_i,a_1,a_2) 
=  8 \alpha_e \alpha_s\ T_R\,\delta^{a_1a_2}
\left[ \left(\sum_{q} Q_q^2\right) \mathrm{M}_{q}(s,t) 
+ Q_t^2 \,\mathrm{M}_{t}(s,t)\right],
\label{eq:ampdef}
\end{align}
where $\alpha_e = e^2/(4 \pi)$ and $Q_{q}$ denotes the electric charge of quark $q$. 
Near the production threshold of an intermediate $t\bar{t}$ pair, $m_{\gamma\gamma} = \sqrt{s} 
\simeq 2 m_t$, we define
$E \equiv m_{\gamma\gamma}-2m_t, \, \beta \equiv \sqrt{1-4m_t^2/\mgg^2+i\delta}$. 
Consequently the scattering angle is given by $\cos\theta = 1 + t\,(1-\beta^2) / (2\,\mt^2)$.  
Close to the threshold, the amplitude $\mathrm{M}_{t}$ can be parametrised 
as~\cite{Melnikov:1994jb,Kawabata:2016aya}
\begin{align}
{\mathrm M}_{t}^{\rm NR} = {\mathcal A}_t(\theta) +
{\mathcal B}_t(\beta)\, G(\vec0;{\mathcal E}) + \mathcal{O}(\beta^3),
\label{eq:thr}
\end{align}
where ${\mathcal E}=E+i\Gamma_t$ includes the top-quark decay width 
$\Gamma_t$\footnote{It has been shown in ref.~\cite{Melnikov:1993np} that in the non-relativistic limit the top width can be consistently included by calculating the cross section for stable top quarks supplemented by such a replacement up to next-to-leading-order according to the NRQCD power counting.}. 
In this parametrisation, the amplitude ${\mathrm M}_{t}^{\rm NR}$ is split into two parts: 
${\mathcal B}_t(\beta)\, G(\vec0;{\mathcal E})$, which contains the $t\bar{t}$ bound-state effects, 
and $ {\mathcal A}_t(\theta)$, which does not. 
The term ${\mathcal B}_t(\beta)\, G(\vec0;{\mathcal E})$ contains the effects from resumming the 
non-relativistic static potential interactions, where the Green's function $G(\vec0;{\mathcal E})$ 
is obtained by solving the non-relativistic Schr\"odinger equation describing a colour-singlet $t\bar{t}$ bound state:
\begin{equation}
\left( -\frac{\nabla^2}{m_t} + V(r) - {\mathcal E} \right) G(\vec{r};{\mathcal E}) =
\delta (\vec{r}),
\label{eq:schroedinger}
\end{equation} 
with the NLO QCD static potential given in~\cite{Fischler:1977yf,Billoire:1979ih,Kawabata:2016aya}.
We remark that the mass $m_t$ appearing in \eqref{eq:schroedinger} is the pole mass of the top quark. 
$G(\vec0;{\mathcal E})$ is the $r \rightarrow 0$ limit of the Green's function $G(\vec{r};{\mathcal E})$.
The real part of the NLO Green's function at $r=0$ is divergent and therefore has to be renormalised. 
We adopt the $\overline{\rm MS}$ scheme, thus introducing a scale $\mu$ into the renormalised Green's 
function~\cite{Beneke:1998rk,Hoang:1998xf,Beneke:1999ff,Hoang:2001mm}. 
~\\

To retain NRQCD resummation effects and, at the same time, keep the matched cross section accurate 
up to NLO in the fixed-order power counting, we \textit{define} our NLO-matched cross section as follows,  
\begin{eqnarray}
\sigma^\mathrm{match}_\mathrm{LO} &\equiv&
a_s^2(\mu_R) \int_{\tau_{min}}^{1} \mathrm{d} \tau \frac{\mathrm{d} \mathcal{L}_{gg}(\mu_F)}{\mathrm{d} \tau} \, \mathcal{N}_{gg}
\int \mathrm{d}\Phi_2 \, \Big| \mathcal{M}_{B} + 
\mathbf{c}\, \left( {\mathcal B}(\mu)\, G(\vec0;{\mathcal E}, \mu) 
- {\mathrm M}_{\rm{OC}}^{(0)} \right) \Big|^{\,2}, \nonumber\\
\nonumber\\
\sigma^{{\rm match}}_\mathrm{NLO} &\equiv& \sigma^\mathrm{match}_\mathrm{LO} \nonumber\\
&+& a_s^3(\mu_R)\, \int_{\tau_{min}}^{1} \mathrm{d} \tau \frac{\mathrm{d} \mathcal{L}_{gg}(\mu_F)}{\mathrm{d} \tau} \, \mathcal{N}_{gg} \,
\int \mathrm{d}\Phi_2 \, 2\,\mathrm{Re} \left[ \mathcal{M}_{B}^{\dagger} 
\left( {\cal M}_V(\mu_R) - \mathbf{c}\,{\mathrm M}_{\rm{OC}}^{(1)}(\mu) \right) \right] \nonumber\\
&+& a_s^3(\mu_R) \int_{\tau_{min}}^{1} \mathrm{d} \tau  \sum_{ij} \frac{\mathrm{d} \mathcal{L}_{ij}(\mu_F)}{\mathrm{d} \tau} \, 
\mathcal{N}_{ij}\, \int \mathrm{d}\Phi_3 \, \Big|{\cal M}_{R,[ij]}(\mu_R) \Big|^{\,2} 
+ \sigma_C \left( \mu_F, \mu_R \right),\, 
\label{eq:matchedXsect}
\end{eqnarray}
where $\mathcal{N}_{ij}$ contains the flux factor and the average over
spins and colours of the initial state partons of flavour $i$ and $j$,
e.g.~$\mathcal{N}_{gg} = \frac{1}{2s} \frac{1}{64} \frac{1}{4}$. 
We have introduced the luminosity factors $\mathcal{L}_{ij}$,
defined as $ \frac{\mathrm{d} \mathcal{L}_{ij}}{\mathrm{d} \tau} \equiv 
\int_{\tau}^{1} \frac{\mathrm{d} x}{x} f_{i} (x,\mu_F) \, f_{j} (\frac{\tau}{x},\mu_F)$,
where $f_{i}(x,\mu_F)$ is the parton distribution function (PDF) of a parton with momentum fraction $x$ and flavour
$i$ (including gluons) and $\mu_F$ is the factorisation scale.
The usual renormalisation scale is denoted by $\mu_R$. 
The 2- and 3-particle phase-space integration measures are denoted by $\mathrm{d}\Phi_2$ and $\mathrm{d}\Phi_3$.
The symbol $\mathbf{c} \equiv 32\,\!\pi \, \alpha_e \, Q_t^2 \, T_R\,\delta^{a_1a_2}$ 
collects constants which have been extracted in the definition of ${\mathrm M}_t$.
$\mathcal{M}_{B}$ and ${\cal M}_V(\mu_R)$ are the one-loop and UV renormalised two-loop amplitudes, 
respectively, with power factors in the coupling $a_s \equiv \alpha_s(\mur)/(4\pi)$ pulled out.
The real-radiation contributions with the factors of $a_s$ extracted are symbolically denoted by ${\cal M}_{R,[ij]}$ 
and the collinear-subtraction counterterm is denoted by $\sigma_C$ for short.
The ${\mathrm M}_{\rm{OC}}^{(0)}$ and ${\mathrm M}_{\rm{OC}}^{(1)}(\mu)$ denote the LO and NLO double-counted part 
of the amplitude in the matching, given by  
\begin{align}
{\mathrm M}_{\mathrm{OC}}^{(0)} & = \mathcal{B}_t^{(0)} G^{(0)}(\vec0;E), \nn \\
{\mathrm M}_{\mathrm{OC}}^{(1)} & = \mathcal{B}_t^{(1)} G^{(0)}(\vec0;E) + \mathcal{B}_t^{(0)} G^{(1)}(\vec0;E).
\end{align}
Note that the explicit dependence of ${\mathrm M}_{\rm{OC}}^{(1)}(\mu)$ on the scale $\mu$ 
stems from the renormalisation of the Green's function $G(\vec0;{ E})$, 
while $\mu_R$ comes from the renormalisation of UV divergences in ${\cal  M}_V(\mu_R)$ and $\mu_F$ 
from initial-state collinear factorisation. 
For the numerical evaluation of eq.~\eqref{eq:matchedXsect}, we expand ${\mathrm M}^{(0)}_{\mathrm{OC}}$ and ${\mathrm M}^{(1)}_{\mathrm{OC}}$ to respectively $\mathcal{O}(\beta^3)$ and $\mathcal{O}(\beta^2)$ (see ref.~\cite{Chen:2019fla} for their explicit expressions).

At the two-loop order, the UV-renormalised amplitude ${\mathrm M}_{t}$ (after IR subtraction) contains a Coulomb 
singularity which is logarithmic in $\beta$. This singularity is, however, subtracted by the expanded term 
${\mathrm M}_{\mathrm{OC}}^{(1)}$, while a resummed description of the Coulomb interactions 
is added back by the term ${\mathcal B}_{t} \, G(\vec0;{\mathcal E})$.
For this purpose, we evaluate the Schr\"odinger equation~\eqref{eq:schroedinger} numerically~\cite{Kiyo:2010jm} 
to obtain $G(\vec0;{\mathcal E})$, where we include $\mathcal{O}(\als)$ corrections to the QCD potential~\cite{Fischler:1977yf,Billoire:1979ih}. 
Unlike the calculation in ref.~\cite{Kawabata:2016aya}, we also include $\mathcal{O}(\als)$ corrections to ${\mathcal B}_t$ as listed above.

\section{Results}
\label{sec:results}

Our numerical results are calculated at a hadronic centre-of-mass energy of $13$\,TeV, using the parton distribution functions
PDF4LHC15{\tt\_}nlo{\tt\_}100~\cite{Butterworth:2015oua,CT14,MMHT14,NNPDF}
interfaced  via
LHAPDF~\cite{Buckley:2014ana}, along with the corresponding value for
$\alpha_s$.  For the
electromagnetic coupling, we use $\alpha = 1/137.03$.
The mass of the top quark is fixed to $m_t=173$\,GeV. The top-quark width is set to zero in
the fixed order calculation, and to $\Gamma_t=1.498$\,GeV in the numerical
evaluation of the Green's function $G(\vec0;{\mathcal E},\mu)$ in accordance with ref.~\cite{Kawabata:2016aya}.
We use the cuts $p_{T,\gamma_1}^{\rm{min}}=40$\,GeV,
$p_{T,\gamma_2}^{\rm{min}}=25$\,GeV and $|\eta_\gamma|\leq 2.5$.
No photon isolation cuts are applied.

The factorisation and renormalisation scale uncertainties are
estimated by varying the scales $\mu_{F}$ and $\mu_{R}$. Unless
specified otherwise, the scale variation bands in figures below 
represent the envelopes of a 7-point scale variation with
$\mu_{R,F}=c_{R,F}\,m_{\gamma\gamma}/2$, where $c_R,c_F\in \{2,1,0.5\}$ and where the
extreme variations $(c_R,c_F)=(2,0.5)$ and $(c_R,c_F)=(0.5,2)$ have been omitted. 
The dependence on the scale $\mu$ introduced by renormalisation of the Green's
function $G(\vec{r};{\mathcal E})$ in our NRQCD matched results is investigated separately.
We have validated the massless NLO cross section by comparison to
MCFM version 9.0~\cite{Campbell:2019dru} and find agreement within the
numerical uncertainties for all scale choices. 
~\\

\begin{figure}[hptb]
\begin{center}
    \includegraphics[width=0.6\textwidth]{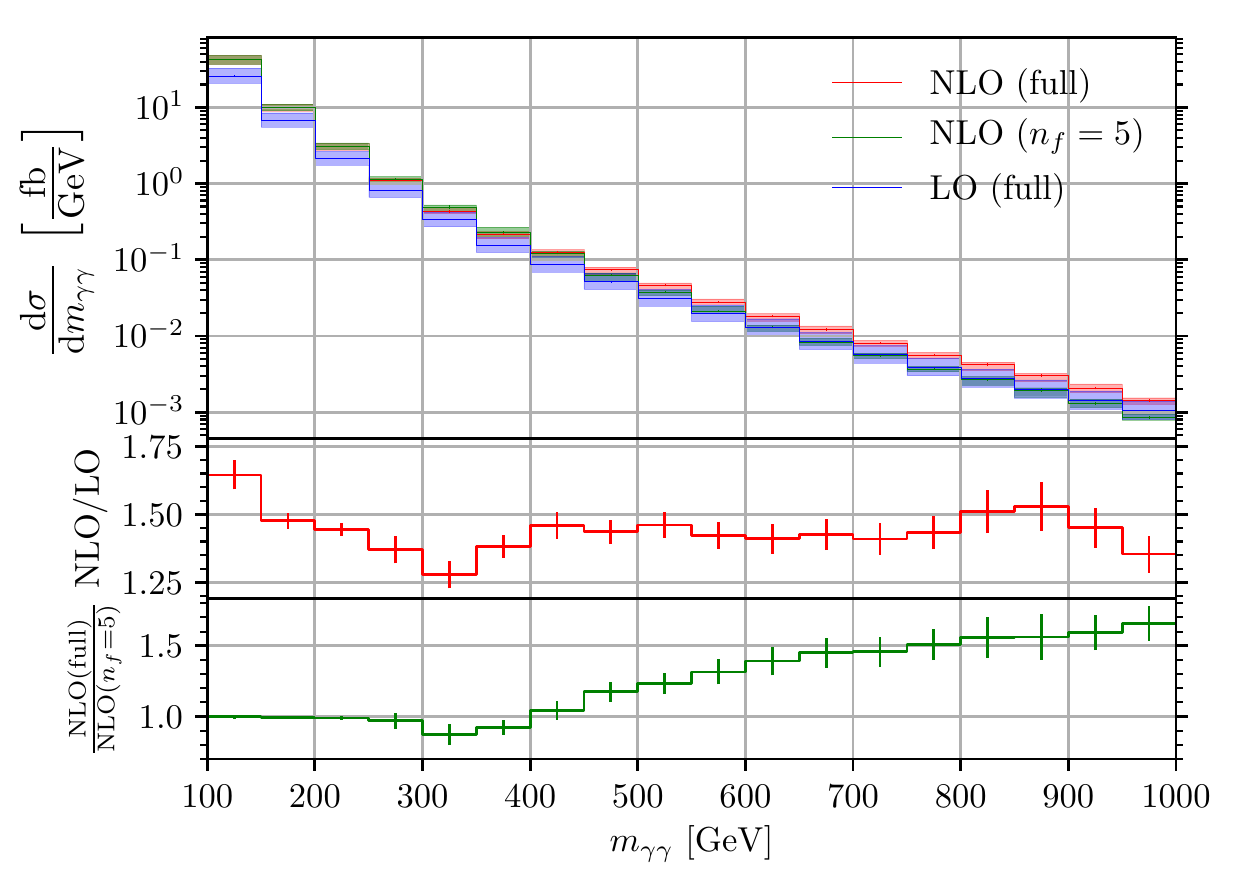}
    \caption{Diphoton invariant mass distribution (fixed order calculation), comparing the result with $n_f=5$ to the result including massive top quark loops.
             The shaded bands show the envelope of the 7-point scale variation as explained in the text.
             The lower panels shows the ratios
             $\mathrm{NLO(full)}/\mathrm{LO(full)}$ and $\mathrm{NLO(full)}/\mathrm{NLO}(n_f=5)$ evaluated at the central scale $\mu_{R}=\mu_{F}=\mgg/2$.
             The bars indicate the uncertainty due to the numerical
             evaluation of the phase-space and loop integrals.}
    \label{fig:myy}
\end{center}
\end{figure}

The distribution of the diphoton invariant mass is shown in fig.~\ref{fig:myy} 
for invariant masses up to 1~TeV, where we show purely fixed order results at LO, 
at NLO with five massless flavours and including top quark loops. 
The ratio plots show the K-factor including the full quark loop content 
and the ratio between the full and the five-flavour NLO cross-section. 
We observe that the scale uncertainties are reduced at NLO, and that
the top quark loops enhance the differential cross section for $\mgg$ values 
far beyond the top quark pair production threshold, asymptotically approaching 
the $n_f=6$ value~\cite{Campbell:2016yrh}.\footnote{We compared against the results shown in~\cite{Maltoni:2018zvp} and find agreement for the central scale choice, while we observe a smaller scale uncertainty band.}

\begin{figure}[hptb]
\begin{center}
    \includegraphics[width=0.6\textwidth]{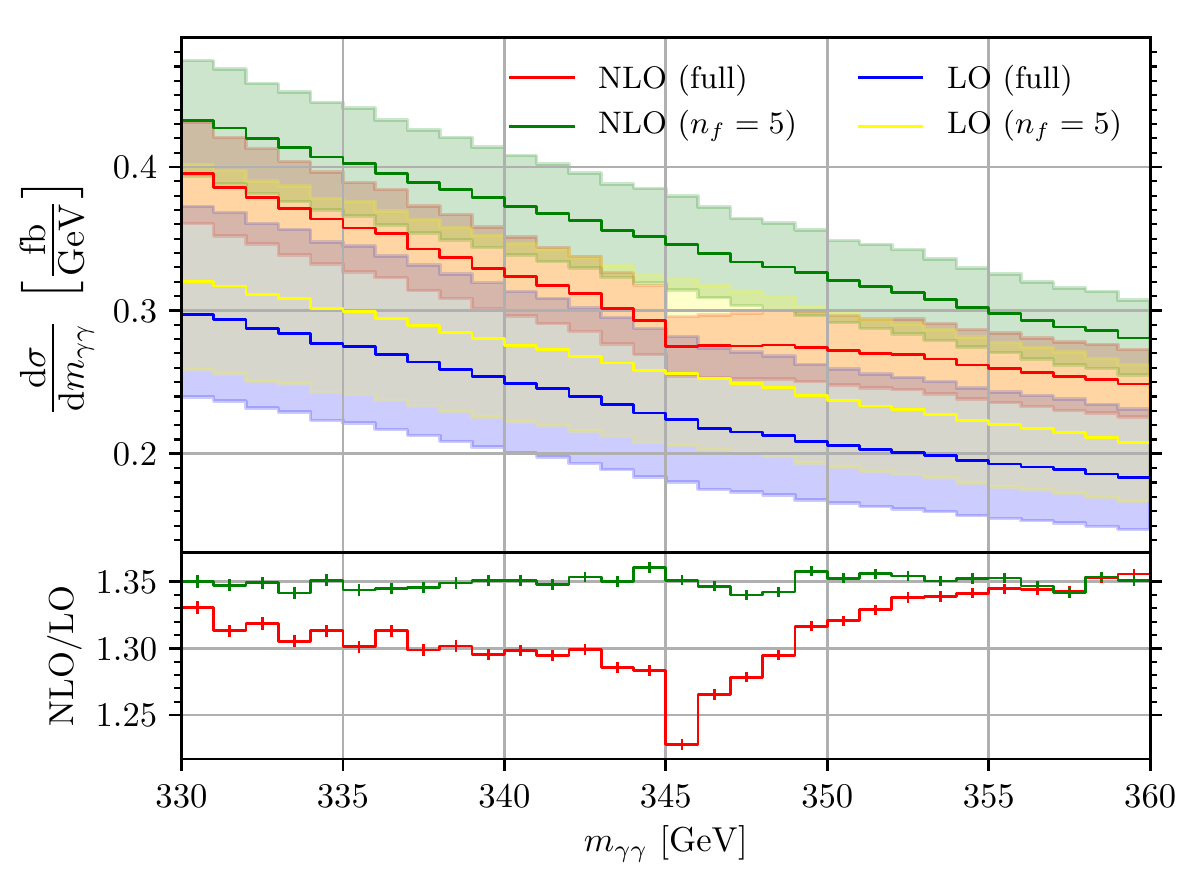}
    \caption{Zoom into the threshold region of the diphoton invariant mass distribution (fixed order calculation), showing the $n_f=5$ and full result separately. The shaded
            bands indicate the scale uncertainties, while the bars indicate uncertainties due to the numerical
            evaluation of the phase-space and loop integrals. The ratio plot in the lower panel shows the ratios $\mathrm{NLO(full)}/\mathrm{LO(full)}$ (red) and
            $\mathrm{NLO(}n_f=5\mathrm{)}/\mathrm{LO(}n_f=5\mathrm{)}$
          (green).}
    \label{fig:myy_toponly_zoom}
\end{center}
\end{figure}

In fig.~\ref{fig:myy_toponly_zoom} 
we zoom into the threshold region, still showing fixed order results only. 
We can clearly see that around the top quark pair production threshold, 
the full result shows a dip and then changes slope, due to the Coulomb effect   
in the virtual amplitude and also the appearance of additional imaginary parts 
above the threshold (see top left diagram of fig.~\ref{fig:2l_diagrams}).

\begin{figure}[hptb]
\begin{center}
    \includegraphics[width=0.8\textwidth]{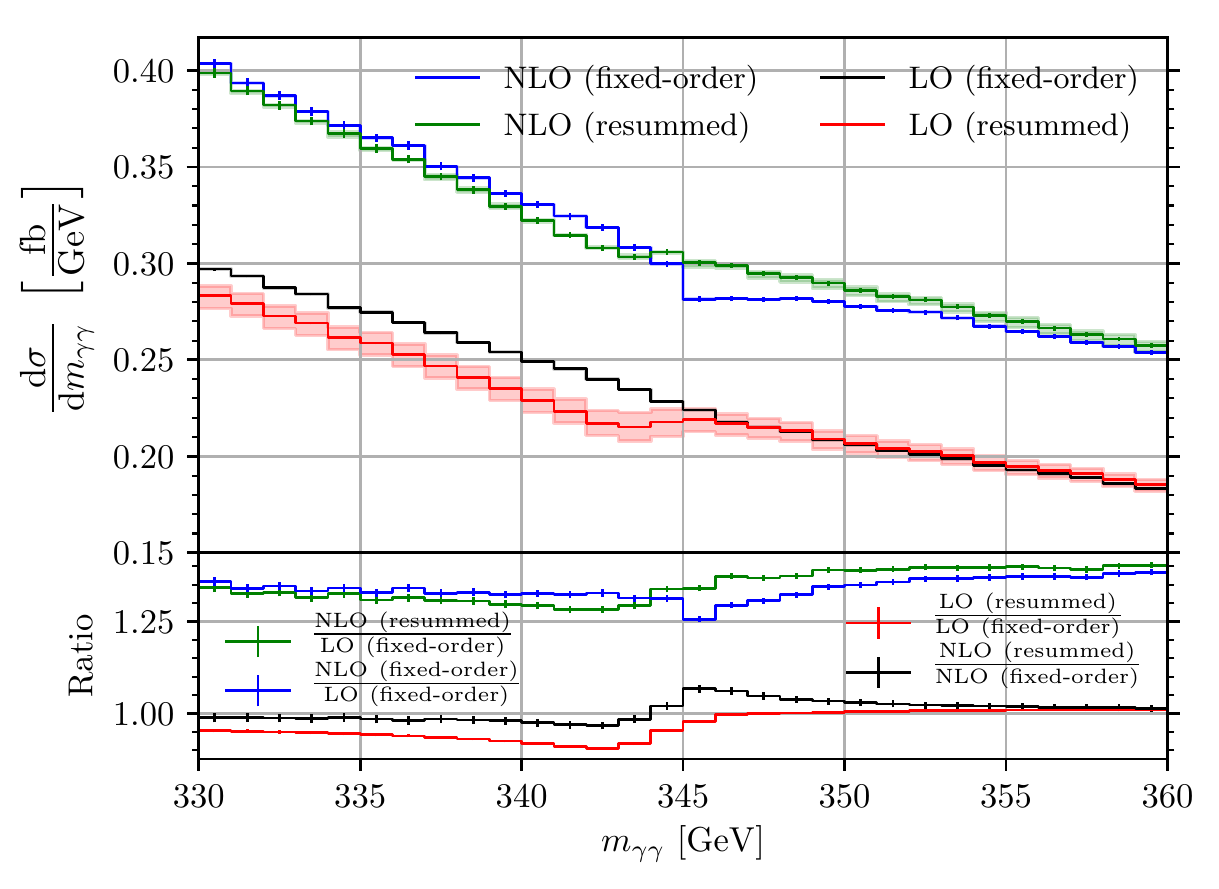}
    \caption{Zoom into the threshold region of the diphoton invariant mass distribution, comparing results with and without NRQCD.
             The shaded bands indicate the scale uncertainty by varying the scale $\mu$ by a factor of 2 around the central scale $\mu=80$\,GeV.
             The renormalisation and the factorisation scales are set to $\mu_R=\mu_F=m_{\gamma\gamma}/2$ and not varied in this plot.
             The bars indicate uncertainties due to the numerical evaluation of the phase-space and loop integrals.}
    \label{fig:myy_NRQCD}
\end{center}
\end{figure}

In fig.~\ref{fig:myy_NRQCD}
we show the $\mgg$-distribution in the threshold region which results from a combination of our fixed order NLO 
(QCD) calculation with the resummation of Coulomb gluon exchanges. 
The scale bands in this figure are produced by varying only $\mu$, the scale associated to the 
renormalisation of the Green's function.
We observe that the dependence on the scale $\mu$ is considerably reduced at NLO compared 
to the leading-order matched cross-section. 
The scale band at NLO is comparable to the size of our numerical uncertainties. 
Further, our leading-order matched cross-section shows a milder dependence on $\mu$ 
than the one presented in ref.~\cite{Kawabata:2016aya}.
This is due to the inclusion of NLO-terms in the coefficient ${\mathcal B}_{t}(\beta)$, 
which was omitted in ref.~\cite{Kawabata:2016aya}.
We find that the ``dip-bump'' structure advertised in ref.~\cite{Kawabata:2016aya} remains 
present at NLO.

\section{Conclusions}

\label{sec:conclusions}
We have calculated diphoton production in gluon fusion at NLO in QCD (i.e.~order $\alpha_s^3$), 
including top quark loop corrections.
Matching our fixed order NLO result to a non-relativistic QCD calculation with resummed Coulomb interactions,   
we obtain an accurate description of the diphoton invariant mass spectrum around the top quark pair production threshold.
In this matched result, we observe a reduction of the renormalisation and factorisation scale uncertainties 
in the threshold region, and an even more drastic reduction of the scale uncertainty related to the renormalised 
NLO Green's function.
These results are promising in view of the possibility to measure the top quark mass from the characteristic 
behaviour of the diphoton invariant mass spectrum around the top quark pair production threshold at a future hadron collider.

\bibliographystyle{JHEP}
\bibliography{refs_ggyy}

\end{document}